\documentclass[]{interact}

\usepackage{epstopdf}
\usepackage[caption=false]{subfig}

\usepackage[numbers,sort&compress]{natbib}
\usepackage{makecell}

\usepackage{hyperref}

\bibpunct[, ]{[}{]}{,}{n}{,}{,}

\theoremstyle{plain}

\theoremstyle{definition}

\theoremstyle{remark}

\begin{document}


\title{
Computer-assisted Pronunciation Training - Speech synthesis is almost all you need
}

\author{
\name{Daniel Korzekwa\textsuperscript{1,2}, Jaime Lorenzo-Trueba\textsuperscript{1}, Thomas Drugman\textsuperscript{1}, Bozena Kostek\textsuperscript{2}}
\affil{\textsuperscript{1}Amazon Speech Research
\\ \textsuperscript{2}Gdansk University of Technology, Faculty of ETI, Poland}
}


\maketitle

\href{https://www.sciencedirect.com/science/article/pii/S0167639322000863?utm_campaign=STMJ_AUTH_SERV_PUBLISHED&utm_medium=email&utm_acid=78131116&SIS_ID=&dgcid=STMJ_AUTH_SERV_PUBLISHED&CMX_ID=&utm_in=DM270343&utm_source=AC_}{Go to the article published in Speech Communication Journal}

\bigskip

\begin{abstract}
The research community has long studied computer-assisted pronunciation training (CAPT) methods in non-native speech. Researchers focused on studying various model architectures, such as Bayesian networks and deep learning methods, as well as on the analysis of different representations of the speech signal. Despite significant progress in recent years, existing CAPT methods are not able to detect pronunciation errors with high accuracy (only 60\% precision at 40\%-80\% recall). One of the key problems is the low availability of mispronounced speech that is needed for the reliable training of pronunciation error detection models. If we had a generative model that could mimic non-native speech and produce any amount of training data, then the task of detecting pronunciation errors would be much easier. We present three innovative techniques based on phoneme-to-phoneme (P2P), text-to-speech (T2S), and speech-to-speech (S2S) conversion to generate correctly pronounced and mispronounced synthetic speech. We show that these techniques not only improve the accuracy of three machine learning models for detecting pronunciation errors but also help establish a new state-of-the-art in the field. Earlier studies have used simple speech generation techniques such as P2P conversion, but only as an additional mechanism to improve the accuracy of pronunciation error detection. We, on the other hand, consider speech generation to be the first-class method of detecting pronunciation errors. The effectiveness of these techniques is assessed in the tasks of detecting pronunciation and lexical stress errors. Non-native English speech corpora of German, Italian, and Polish speakers are used in the evaluations. The best proposed S2S technique improves the accuracy of detecting pronunciation errors in AUC metric by 41\% from 0.528 to 0.749 compared to the state-of-the-art approach.

\end{abstract}


\section{Introduction}

Language plays a key role in online education, giving people access to large amounts of information contained in articles, books, and video lectures. Thanks to spoken language and other forms of communication, such as a sign-language, people can participate in interactive discussions with teachers and take part in lively brainstorming with other people. Unfortunately, education is not available to everybody. According to the UNESCO report, 40\% of the global population do not have access to education in the language they understand \cite{unesco2016if}.  ‘If you don’t understand, how can you learn?' the report says. English is the leading language on the Internet, representing 25.9\% of the world's population \cite{statista2020}. Regrettably, research by EF (Education First) \cite{EPI2020} shows a large disproportion in English proficiency across countries and continents. People from regions of 'very low' language proficiency, such as the Middle East, are unable to navigate through English-based websites or communicate with people from an English-speaking country.

Computer-Assisted Language Learning (CALL) helps to improve the English language proficiency of people in different regions \cite{levy2013call}. CALL relies on computerized self-service  tools that are used by students to practice a language, usually a foreign language, also known as a non-native (L2) language. Students can practice multiple aspects of the language, including grammar, vocabulary, writing, reading, and speaking. Computer-based tools can also be used to measure student's language skills and their learning potential by using Computerized Dynamic Assessment (C-DA) test \cite{mehri2019diagnosing}. CALL can complement traditional language learning provided by teachers. It also has a chance to make second language learning more accessible in scenarios where traditional ways of learning languages are not possible due to the cost of learning or the lack of access to foreign language teachers.

Computer-Assisted Pronunciation Training (CAPT) is a part of CALL responsible for learning pronunciation skills. It has been shown to help people practice and improve their pronunciation skills \cite{neri2008effectiveness,golonka2014technologies, tejedor2020assessing}. CAPT consists of two components: an automated pronunciation evaluation component \cite{leung2019cnn,zhang2021text,korzekwa2021mispronunciation} and a feedback component \cite{ai2015automatic}. The automated pronunciation evaluation component is responsible for detecting pronunciation errors in spoken speech, for example, for detecting words pronounced incorrectly by the speaker. The feedback component informs the speaker about mispronounced words and advises  how to pronounce them correctly. This article is devoted to the topic of automated detection of pronunciation errors in non-native speech. This area of CAPT can take advantage of technological advances in machine learning and bring us closer to creating a fully automated assistant based on artificial intelligence for language learning.

The research community has long studied the automated detection of pronunciation errors in non-native speech. Existing work has focused on various tasks such as detecting mispronounced phonemes \cite{leung2019cnn} and lexical stress errors \cite{ferrer2015classification}. Researcher have given most attention to studying various machine learning models such as Bayesian networks \cite{witt2000phone, Hongyan2011} and deep learning methods \cite{leung2019cnn, zhang2021text}, as well as analyzing different representations of the speech signal such as prosodic features (duration, energy and pitch) \cite{chen2010automatic_2}, and cepstral/spectral features \cite{ferrer2015classification, shahin2016automatic, leung2019cnn}. Despite significant progress in recent years, existing CAPT methods detect pronunciation errors with relatively low accuracy of 60\% precision at 40\%-80\% recall \cite{leung2019cnn,korzekwa2021mispronunciation,zhang2021text}.  Highlighting correctly pronounced words as pronunciation errors by the CAPT tool can demotivate students and lower the confidence in the tool. Likewise, missing pronunciation errors can slow down the learning process.

One of the main challenges with the existing CAPT methods is poor availability of mispronounced speech, which is required for the reliable training of pronunciation error detection models. We propose a reformulation of the problem of pronunciation error detection as a task of synthetic speech generation. Intuitively, if we had a generative model that could mimic mispronounced speech and produce any amount of training data, then the task of detecting pronunciation errors would be much easier. The probability of pronunciation errors for all the words in a sentence can then be calculated using the Bayes rule \cite{bishop2006pattern}. In this new formulation, we move the complexity to learning the speech generation process that is well suited to the problem of limited speech availability \cite{huybrechts2021low, shah21_ssw, fazel21_interspeech}. The proposed method outperforms the state-of-the-art model \cite{leung2019cnn} in detecting pronunciation errors in AUC metric by 41\% from 0.528 to 0.749 on the GUT Isle Corpus of L2 Polish speakers. 

To put the new formulation of the problem into action, we propose three innovative techniques based on phoneme-to-phoneme (P2P), text-to-speech (T2S), and speech-to-speech (S2S) conversion to generate correctly pronounced and mispronounced synthetic speech. We show that these techniques not only improve the accuracy of three machine learning models for detecting pronunciation errors but also help establish a new state-of-the-art in the field. The effectiveness of these techniques is assessed in two tasks: detecting mispronounced words (replacing, adding, removing phonemes, or pronouncing an unknown speech sound) and detecting lexical stress errors. The results presented in this study are the culmination of our recent work on speech generation in pronunciation error detection task \cite{korzekwa2021mispronunciation,korzekwa21b_interspeech,korzekwa21_interspeech}, including a new S2S technique.

In short, the contributions of the paper are as follows:
\begin{itemize}
\item A new paradigm for the automated detection of pronunciation errors is proposed, reformulating the problem as a task of generating synthetic speech.
\item A unified probabilistic view on P2P, T2S, and S2S techniques is presented in the context of detecting pronunciation errors.
\item A new S2S method to generate synthetic speech is proposed, which outperforms the state-of-the-art model \cite{leung2019cnn} in detecting pronunciation errors.
\item Comprehensive experiments are described to demonstrate the effectiveness of speech generation in the tasks of pronunciation and lexical stress error detection.
\end{itemize}

The outline of the rest of this paper is: Section \ref{sec:related_work} presents related work. Section \ref{sec:synth_speech_generation_methods} describes the proposed methods of generating synthetic speech for automatic detection of pronunciation errors. Section \ref{sec:human_speech_corpora} describes the human speech corpora used to train the pronunciation error detection models in the experiments. Section \ref{sec:experiments_pronunciation} presents experiments demonstrating the effectiveness of various synthetic speech generation methods in improving the accuracy of the detection of pronunciation and lexical stress errors. Finally, conclusions and future work are presented in Section \ref{sec:conclusions}.

\section{Related work}
\label{sec:related_work}

\subsection{Pronunciation error detection}

\subsubsection{Phoneme recognition approaches} 

Most existing CAPT methods are designed to recognize the phonemes pronounced by the speaker and compare them with the expected (canonical) pronunciation of correctly pronounced speech \cite{witt2000phone,li2016mispronunciation,sudhakara2019improved,leung2019cnn}. Any discrepancy between the recognized and canonical phonemes results in a pronunciation error at the phoneme level. Phoneme recognition approaches generally fall into two categories: methods that align a speech signal with phonemes (forced-alignment techniques) and methods that first recognize the phonemes in the speech signal and then align the recognized and canonical phoneme sequences. Aside these two categories, CAPT methods can be split into multiple other categories:

Forced-alignment techniques \cite{Hongyan2011, li2016mispronunciation, sudhakara2019improved, cheng2020asr} are based on the work of Franco et al. \cite{franco1997automatic} and the Goodness of Pronunciation (GoP) method \cite{witt2000phone}. In the first step, GoP uses Bayesian inference to find the most likely alignment between canonical phonemes and the corresponding audio signal (forced alignment).  In the next step, GoP calculates the ratio between the likelihoods of the canonical and the most likely pronounced phonemes. Finally, it detects mispronunciation if the ratio drops below a certain threshold.  GoP has been further extended with Deep Neural Networks (DNNs), replacing the Hidden Markov Model (HMM) and Gaussian Mixture Model (GMM) techniques for acoustic modeling \cite{li2016mispronunciation,sudhakara2019improved}. Cheng et al. \cite{cheng2020asr} improves GoP performance with the hidden representation of speech extracted in an unsupervised way. This model can detect pronunciation errors based on the input speech signal and the reference canonical speech signal, without using any linguistic information such as text and phonemes. 

The methods that do not use forced-alignment recognize the phonemes pronounced by the speaker purely from the speech signal and only then align them with the canonical phonemes \cite{Minematsu2004PronunciationAB, harrison2009implementation, Lee2013PronunciationAV, plantinga2019towards,Sudhakara2019NoiseRG,zhang2020end}. Leung et al. \cite{leung2019cnn} use a phoneme recognizer that recognizes phonemes only from the speech signal. The phoneme recognizer is based on Convolutional Neural Network (CNN), a Gated Recurrent Unit (GRU), and Connectionist Temporal Classification (CTC) loss. Leung et al. report that it outperforms other forced-alignment \cite{li2016mispronunciation} and forced-alignment-free \cite{harrison2009implementation} techniques in the task of detecting mispronunciations at the phoneme-level in L2 English. 

There are two challenges with presented approaches for pronunciation error detection. First, phonemes pronounced by the speaker must be recognized accurately, which has been proved difficult \cite{zhang2021text, chorowski2014end, chorowski2015attention, bahdanau2016end}. Phoneme recognition is difficult, especially in non-native speech, as different languages have different phoneme spaces. Second, standard approaches assume only one canonical pronunciation of a given text, but this assumption is not always true due to the phonetic variability of speech, e.g., differences between regional accents. For example, the word `enough' can be pronounced by native speakers in multiple ways:  /ih n ah f/ or /ax n ah f/ (short `i' or `schwa' phoneme at the beginning). In our previous work, we solve these problems by creating a native speech pronunciation model that returns the probability of the sentence to be spoken by a native speaker \cite{korzekwa2021mispronunciation}.

Techniques based on phoneme recognition can be supplemented by a reference speech signal obtained  from the speech database \cite{xiao2018paired, nicolao2015automatic, wang2019child} or generated from the phonetic representation \cite{korzekwa2021mispronunciation, qian2010capturing}. Xiao et al. \cite{xiao2018paired} use a pair of speech signals from a student and a native speaker to classify native and non-native speech. Mauro et al. \cite{nicolao2015automatic} use the speech of the reference speaker to detect mispronunciation errors at the phoneme level. Wang et al. \cite{wang2019child} use Siamese networks to model the discrepancy between normal and distorted children's speech. Qian et al. \cite{qian2010capturing} propose a statistical  model of pronunciation in which they build a model that generates hypotheses of mispronounced speech. 

In this work, we use the end-to-end method to detect pronunciation errors directly, without having to recognize phonemes as an intermediate step. The end-to-end approach is discussed in more detail in the next section.

\subsubsection{End-to-end methods} 

The phoneme recognition approaches presented so far rely on phonetically transcribed speech labeled by human listeners. Phonetic transcriptions are needed to train a phoneme recognition model. Human-based transcription is a time-consuming task, especially with L2 speech, where listeners need to recognize mispronunciation errors. Sometimes L2 speech transcription may be even impossible because different languages have different phoneme sets, and it is unclear which phonemes were pronounced by the speaker. In our recent work, we have introduced a novel model (known as WEAKLY-S,  i.e., weakly supervised) for detecting pronunciation errors at the world level that does not require phonetically transcribed L2 speech \cite{korzekwa21b_interspeech}. During training, the model is weakly supervised, in the sense that in L2 speech, only mispronounced words are marked, and the data do not need to be phonetically transcribed. In addition to the primary task of detecting mispronunciation errors at the world level, the second task uses a phoneme recognizer trained on automatically transcribed L1 speech. 

Zhang et al. \cite{zhang2021text} employ a multi-task model with two tasks: phoneme-recognition and pronunciation error detection tasks. Unlike our WEAKLY-S model, they use the Needleman-Wunsch algorithm \cite{needleman1970general} from  bioinformatics to align the canonical and recognized phoneme sequences, but this algorithm cannot be tuned to detect pronunciation errors. The WEAKLY-S model automatically learns the alignment, thus eliminating a potential source of inaccuracy. The alignment is learned through an attention mechanism that automatically maps the speech signal to a sequence of pronunciation errors at the word level. Tong et al. [39] propose to use a multi-task framework in which a neural network model is used to learn the joint space between the acoustic characteristics of adults and children. Additionally, Duan et al. \cite{duan2019cross} propose a multi-task model for acoustical modeling with two tasks for native and non-native speech respectively.

The work of Zhang et al. \cite{zhang2021text} and our recent work \cite{korzekwa21b_interspeech} are end-to-end methods of direct estimation of pronunciation errors, setting up a new trend in the field of automated pronunciation assessment. In this article, we use the end-to-end method as well, but we extend it by the S2S method of generating mispronounced speech. 

\subsubsection{Other trends} 

All the works presented so far treat pronunciation errors as discrete categories, at best producing the probability of mispronunciation. In  contrast, Bi-Cheng et al. \cite{yan20_interspeech} propose a model capable of identifying phoneme distortions, giving the user more detailed feedback on mispronunciation. In our recent work, we provide more fine-grained feedback by indicating the severity level of mispronunciation \cite{korzekwa21b_interspeech}.

Active research is conducted not only on modelling techniques but also on speech representation. Xu et al. \cite{xu21k_interspeech} and Peng et al. \cite{peng21e_interspeech} use the Wav2vec 2.0 speech representation that is created in an unsupervised way. They report that it outperforms existing methods and requires three times less speech training data. Lin et al. \cite{lin21j_interspeech} use transfer learning by taking advantage of deep latent features extracted from the Automated Speech Recognition (ASR) acoustic model and report improvements over the classic GOP-based method.

In this work, we use a mel-spectrogram as a speech representation in the pronunciation error detection model. We also use a mel-spectrogram to represent the speech signal in the T2S and S2S methods of generating mispronounced speech.

\subsection{Lexical stress error detection} 

CAPT usually focuses on practicing the pronunciation of phonemes \cite{witt2000phone, leung2019cnn, korzekwa2021mispronunciation}. However, there is evidence that practicing lexical stress improves the intelligibility of non-native English speech \cite{field2005intelligibility,lepage2014intelligibility}. Lexical stress is a phonological feature of a syllable. It is part of the phonological rules that govern how words should be pronounced in a given language. Stressed syllables are usually longer, louder, and expressed with a higher pitch than their unstressed counterparts \cite{jung2018acoustic}. The lexical stress is related to the phonemic representation. For example, placing lexical stress on a different syllable of a word can lead to various phonemic realizations known as `vowel reduction' \cite{bergem1991acoustic}. Students should be able to practice both pronunciation and lexical stress in spoken language. We study both topics to better understand the potential of using speech generation methods in CAPT.

The existing works focus on the supervised classification of lexical stress using Neural Networks \cite{li2018automatic, shahin2016automatic}, Support Vector Machines \cite{chen2010automatic_2, zhao2011automatic}, and Fisher’s linear
discriminant \cite{chen2007using}. There are two popular variants: a) discriminating syllables between primary stress/no stress \cite{ferrer2015classification}, and b) classifying between primary stress/secondary stress/no stress \cite{li2013lexical,li2018automatic}. Ramanathi et al. \cite{ramanathi2019asr} have followed an alternative unsupervised way of classifying lexical stress, which is based on computing the likelihood of an acoustic signal for a number of possible lexical stress representations of a word.  

Accuracy is the most commonly used performance metric, and it indicates the ratio of correctly classified stress patterns on a syllable \cite{li2013lexical} or word level \cite{chen2010automatic_2}. On the contrary, Ferrer et al. \cite{ferrer2015classification}, analyzed the precision and recall metrics to detect lexical stress errors and not just classify them.

Most existing approaches for the classification and detection of lexical stress errors are based on carefully designed features. They start with aligning a speech signal with phonetic transcription, performed via forced-alignment \cite{shahin2016automatic, chen2010automatic_2}. Alternatively, ASR can provide both phonetic transcription and its alignment with a speech signal \cite{li2013lexical}. Then, prosodic features such as duration, energy and pitch \cite{chen2010automatic_2} and cepstral features such as Mel Frequency Cepstral Coefficients (MFCC) and Mel-Spectrogram \cite{ferrer2015classification,shahin2016automatic} are extracted. These features can be extracted on the syllable \cite{shahin2016automatic} or syllable nucleus \cite{ferrer2015classification,chen2010automatic_2} level. Shahin et al. \cite{shahin2016automatic} computes features of neighboring vowels, and Li et al. \cite{li2013lexical} includes the features for two preceding and two following syllables in the model. The features are often preprocessed and normalized to avoid potential confounding variables \cite{ferrer2015classification}, and to achieve better model generalization by normalizing the duration and pitch on a word level \cite{ferrer2015classification,chen2007using}. Li et al. \cite{li2018automatic} adds canonical lexical stress to input features, which improves the accuracy of the model. 

In our recent work, we use attention mechanisms to automatically derive areas of the audio signal that are important for the detection of lexical stress errors  \cite{korzekwa21_interspeech}. In this work, we use the T2S method to generate synthetic lexical stress errors to improve the accuracy of detecting lexical stress errors.

\subsection{Synthetic speech generation for pronunciation error detection}

Existing synthetic speech generation techniques for detecting pronunciation errors can be divided into two categories: data augmentation and data generation.  

Data augmentation techniques are designed to generate new training examples for existing mispronunciation labels. Badenhorst et al. \cite{badenhorst2017limitations} simulate new speakers by adjusting the speed of raw audio signals. Eklund \cite {eklund2019data} generates additional training data by adding background noise and convolving the audio signal with the impulse responses of the microphone of a mobile device and a room.

Data generation techniques are designed to generate new training data with new labels of both correctly pronounced and mispronounced speech. Most existing works are based on the P2P technique to generate mispronounced speech by perturbing the phoneme sequence of the corresponding audio using a variety of strategies \cite{lee2016language, komatsu2019speech, fu2021full, yan2021maximum,korzekwa2021mispronunciation}. In addition to P2P techniques, in our recent work, we use T2S to generate synthetic lexical stress errors \cite{korzekwa21b_interspeech}. Qian et al. \cite{qian2010capturing} introduce a generative model to create hypotheses of mispronounced speech and use it as a reference speech signal to detect pronunciation errors. Recently, we proposed a similar technique to create a pronunciation model of native speech to account for many ways of correctly pronouncing a sentence by a native speaker \cite{korzekwa2021mispronunciation}.

Synthetic speech generation techniques have recently gained attention in other related fields. Fazel et al. \cite{fazel21_interspeech} use synthetic speech generated with T2S to improve accuracy in ASR. Huang et al. \cite{huang2016machine} use a machine translation technique to generate text to train an ASR language model in a low-resource language. At the same time, Shah et al. \cite{shah21_ssw} and Huybrechts et al. \cite{huybrechts2021low} employ S2S voice conversion to improve the quality of speech synthesis in the data reduction scenario.

All the presented works on the detection of pronunciation errors treat synthetic speech generation as a secondary contribution. In this article, we present a unified perspective of synthetic speech generation methods for detecting pronunciation errors. This article extends our previous work \cite{korzekwa2021mispronunciation,korzekwa21_interspeech,korzekwa21b_interspeech} and introduces a new S2S method to detect pronunciation errors. To the best of our knowledge, there are no papers devoted to generating pronunciation errors with the S2S technique and using it in the detection of pronunciation errors.

\section{Methods of generating pronunciation errors}
\label{sec:synth_speech_generation_methods}

To detect pronunciation errors, first, the spoken language must be separated from other factors in the signal and then incorrectly pronounced speech sounds have to be identified. Separating speech into multiple factors is difficult, as speech is a complex signal. It consists of prosody (F0, duration, energy), timbre of the voice, and the representation of the spoken language. Spoken language is defined by the sounds (phones) perceived by people. Phones are the realizations of phonemes - a human abstract representation of how to pronounce a word/sentence. Speech may also present variability due to the recording channel and environmental effects such as noise and reverberation. Detecting pronunciation errors is very challenging, also because of the limited amount of recordings with mispronounced speech. To address these challenges, we reformulate the problem of pronunciation error detection as the task of synthetic speech generation.

Let $\mathbf{s}$ be the speech signal, $\mathbf{r}$ be the sequence of phonemes that the user is trying to pronounce (canonical pronunciation), and $\mathbf{e}$ be the sequence of probabilities of mispronunciation at the phoneme or word level. The original task of detecting pronunciation errors is defined by:

\begin{equation}
 \mathbf{e} \sim p(\mathbf{e}|\mathbf{s},\mathbf{r})
\end{equation}

where the formulation of the problem as the task of synthetic speech generation is defined as follows:

\begin{equation}
 \mathbf{s} \sim p(\mathbf{s}|\mathbf{e},\mathbf{r})
\end{equation} 

The probability of pronunciation errors for all the words in a sentence can then be calculated using the Bayes rule \cite{bishop2006pattern}:

\begin{equation}
p(\mathbf{e}|\mathbf{s},\mathbf{r}) = \frac{p(\mathbf{e}|\mathbf{r})p(\mathbf{s}|\mathbf{e},\mathbf{r})}{p(\mathbf{s}|\mathbf{r})}
\label{eqn:pron_error_detection_bayes_rule}
\end{equation} From Equation \ref{eqn:pron_error_detection_bayes_rule}, one can see that there is no need to directly learn the probability of pronunciation errors $p(\mathbf{e}|\mathbf{s},\mathbf{r})$, since the complexity of the problem has now been transferred to learning the speech generation process $p(\mathbf{s}|\mathbf{e}, \mathbf{r})$. Such a formulation of the problem opens the way to the inclusion of additional prior knowledge into the model:

\begin{enumerate}
\item Replacing the phoneme in a word while preserving the original speech signal results in a pronunciation error (P2P method).
\item Changing the speech signal while retaining the original pronunciation results in a pronunciation error (T2S method).
\item There are many variations of mispronounced speech that differ in terms of the voice timbre and the prosodic aspects of speech (S2S method).
\end{enumerate}

To solve Equation \ref{eqn:pron_error_detection_bayes_rule}, we use Markov Chain Monte Carlo Sampling (MCMC) \cite{koller2009probabilistic}. In this way, the prior knowledge can be incorporated by generating $N$ training examples $\{\mathbf{e_i},\mathbf{s_i}, \mathbf{r_i}\}$ for $i=1..N$ with the use of P2P (prior knowledge 1), T2S (prior knowledge 2), and S2S (prior knowledge 3) methods. Accounting for the prior knowledge, intuitively corresponds to an increase in the amount of training data, which contributes to outperforming state-of-the-art models for detecting pronunciation errors, as presented in Section \ref{sec:experiments_pronunciation}. Equation \ref{eqn:pron_error_detection_bayes_rule} can then be optimized with standard gradient-based optimization techniques. In the following subsections, we present the P2P conversion, T2S, and S2S methods of generating correctly and incorrectly pronounced speech in details. 

\subsection{P2P method}
\label{sec:p2p_method}

\begin{figure}
\centering
\includegraphics[height=3.8cm]{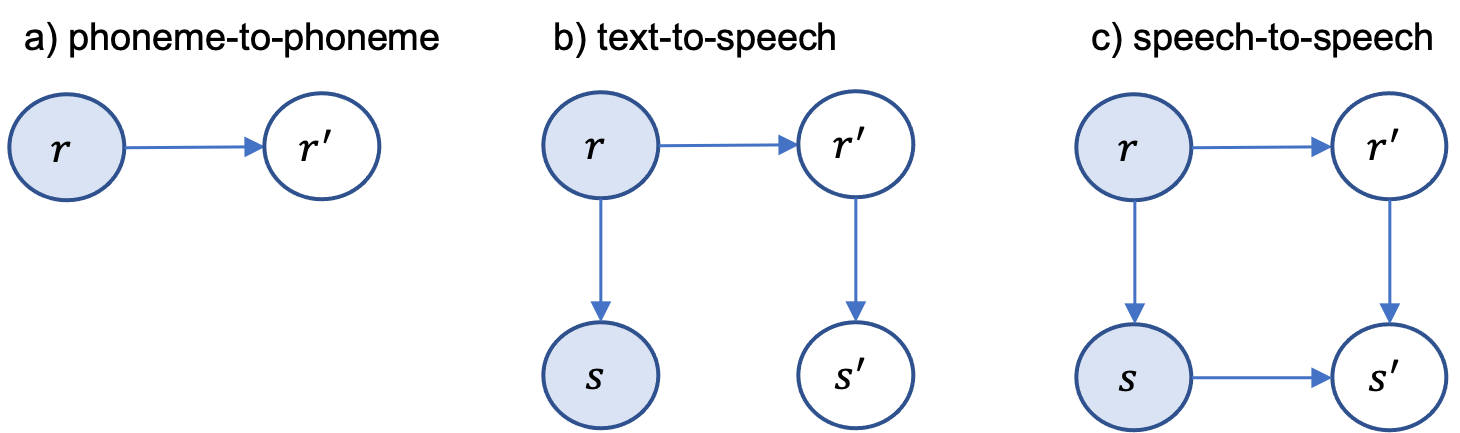}
\caption{Probabilistic graphical models for three methods to generate pronunciation errors: P2P, T2S and S2S. Empty circles represent hidden (latent) variables, while filled (blue) circles represent observed variables.  $\mathbf{s}$ - the speech signal, $\mathbf{r}$ - the sequence of phonemes that the user is trying to pronounce (canonical pronunciation), the superscript $\mathbf{'}$ represents a variable with generated mispronunciations.} 
\label{fig:pgms}
\end{figure}

To generate synthetic mispronounced speech, it is enough to start with correctly pronounced speech and modify the corresponding sequence of phonemes. This simple idea does not even require generating the speech signal itself. It can be observed that the probability of mispronunciations depends on the discrepancy between the speech signal and the corresponding canonical pronunciation. This leads to the P2P conversion model shown in Figure \ref{fig:pgms}a.

Let $\{\mathbf{e_{noerr}}, \mathbf{s},\mathbf{r}\}$ be a single training example containing: the sequence of $0s$ denoting correctly pronounced phonemes, the speech signal, and the sequence of phonemes representing the canonical pronunciation.  Let $\mathbf{r^{'}}$ be the sequence of phonemes with injected mispronunciations such as phoneme replacements, insertions, and deletions:
\begin{equation}
\mathbf{r^{'}} \sim p(\mathbf{r^{'}}|\mathbf{r})
\end{equation}
then the probability of mispronunciation for the $j^{th}$ phoneme is defined by:

\begin{equation}
e_j^{'} = 
\left\{ 
  \begin{array}{rl}
     1 &\mbox{ if $r_j^{'} != r_j $} \\
     0 &\mbox{ otherwise}
   \end{array} 
 \right.
 \label{eqn:ptp_pron_error}
\end{equation}
The probabilities of mispronunciation can be projected from the level of phonemes to the level of words. A word is treated as mispronounced if at least one pair of phonemes in the word $\{r_j^{'},r_j\}$ does not match. At the end of this process, a new training example is created with artificially introduced pronunciation errors: $\{\mathbf{e_{err}}, \mathbf{s},\mathbf{r^{'}}\}$. Note that the speech signal $\mathbf{s}$ in the new training example is unchanged from the original training example, and only phoneme transcription is manipulated.

\bigskip

\textbf{Implementation}

\bigskip

To generate synthetic pronunciation errors, we use a simple approach of perturbing phonetic transcription for the corresponding speech audio.  First, we sample these utterances with replacement from the input corpora of human speech. Then, for each utterance, we replace the phonemes with random phonemes with a given probability. 

\subsection{T2S method}
\label{sec:t2s_method}

The T2S method expands on P2P by making it possible to create speech signals that match the synthetic mispronunciations. The T2S method for generating mispronounced speech is a generalization of the P2P method, as can be seen by the comparison of the two methods shown in Figures \ref{fig:pgms}a and \ref{fig:pgms}b.

One problem with the P2P method is that it cannot generate a speech signal for the newly created sequence of phonemes $\mathbf{r^{'}}$. As a result, pronunciation errors will dominate in the training data containing new sequences of phonemes $\mathbf{r^{'}}$. Therefore, it will be possible to detect pronunciation errors only from the canonical representation $\mathbf{r^{'}}$, ignoring information contained in the speech signal. To mitigate this issue, there should be two training examples for the phonemes $\mathbf{r^{'}}$, one representing mispronounced speech: $\{\mathbf{e_{err}}, \mathbf{s},\mathbf{r^{'}}\}$, and the second one for correct pronunciation: $\{\mathbf{e_{noerr}}, \mathbf{s^{'}},\mathbf{r^{'}}\}$, where: 
\begin{equation}
 \mathbf{s^{'}} \sim p(\mathbf{s^{'}}|\mathbf{e_{noerr}},\mathbf{r^{'}})
\end{equation}

Because we now have the speech signal $\mathbf{s^{'}}$, another training example can be created as: $\{\mathbf{e_{err}}, \mathbf{s^{'}},\mathbf{r}\}$. In summary, T2S method extends a single training example of correctly pronounced speech to four combinations of correctly and incorrect pronunciations: 
\begin{itemize}
\item $\{\mathbf{e_{noerr}}, \mathbf{s},\mathbf{r}\}$ -- correctly pronounced input speech

\item $\{\mathbf{e_{err}}, \mathbf{s},\mathbf{r^{'}}\}$ -- mispronounced speech generated by the P2P method

\item $\{\mathbf{e_{noerr}}, \mathbf{s^{'}},\mathbf{r^{'}}\}$ -- correctly pronounced speech generated by the T2S method

\item $\{\mathbf{e_{err}}, \mathbf{s^{'}},\mathbf{r}\}$ --  mispronounced speech generated by the T2S method
\end{itemize}

\bigskip

\textbf{Implementation}

\bigskip

The synthetic speech is generated with the Neural TTS described by Latorre et al. \cite{latorre2019effect}. The Neural TTS consists of two modules. The context-generation module is an attention-based encoder-decoder neural network that generates a mel-spectrogram from a sequence of phonemes. The Neural Vocoder then converts it into a speech signal. The Neural Vocoder is a neural network of architecture similar to Parallel Wavenet \cite{oord2018parallel}. The Neural TTS is trained using the speech of a single native speaker. To generate words with different lexical stress patterns, we modify the lexical stress markers associated with the vowels in the phonetic transcription of the word. For example, with the input of /r iy1 m ay0 n d/ we can place lexical stress on the first syllable of the word `remind'.

\subsection{S2S method}
\label{sec:s2s_method}

The S2S method is designed to simulate the diverse nature of speech, as there are many ways to correctly pronounce a sentence. The prosodic aspects of speech, such as pitch, duration, and energy, can vary. Similarly, phonemes can be pronounced differently. To mimic human speech, speech generation techniques should allow a similar level of variability. The T2S method outlined in the previous section always produces the same output for the same phoneme input sequence. The S2S method is designed to overcome this limitation.

S2S converts the input speech signal $\mathbf{s}$ in a way to change the pronounced phonemes (phoneme replacements, insertions, and deletions) from the input phonemes $\mathbf{r}$ to target phonemes $\mathbf{r^{'}}$ while preserving other aspects of speech, including voice timbre and prosody (Equation \ref{eqn:sts_equation_for_s} and Figure \ref{fig:pgms}c). In this way, the natural variability of human speech is preserved, resulting in generating many variations of incorrectly pronounced speech. The prosody will differ in various versions of the sentence of the same speaker, while the same sentence spoken by many speakers will differ in the voice timbre. 
\begin{equation}
 \mathbf{s^{'}} \sim p(\mathbf{s^{'}}|\mathbf{e_{noerr}},\mathbf{r^{'}}, \mathbf{s})
 \label{eqn:sts_equation_for_s}
\end{equation}
Similarly to the T2S method, the S2S method outputs four types of speech pronounced correctly and incorrectly: $\{\mathbf{e_{noerr}}, \mathbf{s},\mathbf{r}\}$, $\{\mathbf{e_{err}}, \mathbf{s},\mathbf{r^{'}}\}$, $\{\mathbf{e_{noerr}}, \mathbf{s^{'}},\mathbf{r^{'}}\}$, and $\{\mathbf{e_{err}}, \mathbf{s^{'}},\mathbf{r}\}$.

\bigskip

\textbf{Implementation}

\bigskip

\begin{figure}
\centering
\includegraphics[height=2.7cm]{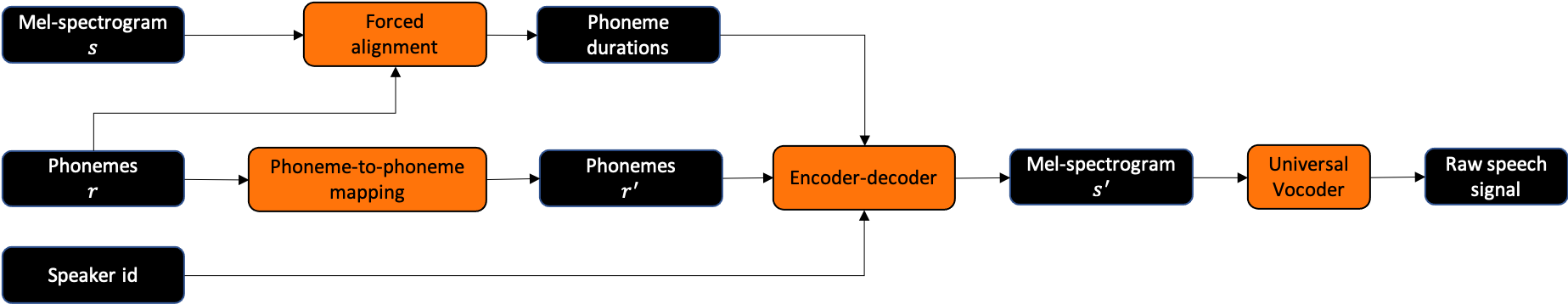}
\caption{Architecture of the S2S model to generate mispronounced synthetic speech while maintaining prosody and voice timbre of the input speech. The black rectangles represent the data (tensors) and the orange boxes represent processing blocks. This color notation is used in all  machine learning model diagrams throughout the article.} 
\label{fig:speech_to_speech_architecture}
\end{figure}

Synthetic speech is generated by introducing mispronunciations into the input speech, while preserving the duration of the phonemes and timbre of the voice. The architecture of the S2S model is shown in Figure \ref{fig:speech_to_speech_architecture}. The mel-spectrogram of the input speech signal $\mathbf{s}$ is forced-aligned with the corresponding canonical phonemes $\mathbf{r}$ to get the duration of the phonemes. The speaker id has to be provided together with the input speech to enable the source speaker's voice to be maintained. Mispronunciations are introduced into the canonical phonemes $\mathbf{r}$ according to the P2P method described in Section \ref{sec:p2p_method}. Mispronounced phonemes $\mathbf{r^{'}}$ along with phonemes duration and speaker id are processed by the encoder-decoder, which generates the mel-spectrogram $\mathbf{s^{'}}$. The encoder-decoder transforms the phoneme-level representation into frame-level features and then generates all mel-spectrogram frames in parallel. The mel-spectrogram is converted to an audio signal with Universal Vocoder \cite{jiao2021universal}. Without the Universal Vocoder, it would not be possible to generate the raw audio signal for hundreds of speakers included in the LibriTTS corpus. Details of the S2S method are shown in the works of Shah et al. \cite{shah21_ssw} and Jiao et al. \cite{jiao2021universal}. The main difference between these two models and our S2S model is the use of the P2P mapping to introduce pronunciation errors.

\subsection{Summary of mispronounced speech generation}

Generation of synthetic mispronounced speech and detection of pronunciation errors were presented from the probabilistic perspective of the Bayes-rule. With this formulation, we can better understand the relationship between P2P, T2S and S2S methods, and see that the S2S method generalizes two simpler methods. Following this reasoning, we can argue that using the Bayes rule gives us a nice mathematical framework to potentially further generalize the S2S method, e.g. by adding a language variable to the model to support multilingual pronunciation error detection. There is another advantage of modelling pronunciation error detection from the probabilistic perspective - it paves the way for joint training of mispronounced speech generation and pronunciation error detection models. In the present work, we are training separate machine learning models for both tasks, but it should be possible to train both models jointly using the framework of Variational Inference \cite{jordan1999introduction} instead of MCMC to infer the probability of mispronunciation in Equation \ref{eqn:pron_error_detection_bayes_rule}.

\section{Speech corpora}
\label{sec:human_speech_corpora}

\subsection{Corpora of continuous speech}

Speech corpora of recorded sentences is a combination of L1 and L2 English speech. L1 speech is obtained from the TIMIT \cite{garofolo1993darpa} and the LibriTTS \cite{Zen2019} corpora. L2 speech comes from the Isle \cite{atwell2003isle} corpus (German and Italian speakers) and the GUT Isle \cite{Weber2020} corpus (Polish speakers). In total, we used 125.28 hours of L1 and L2 English speech from 983 speakers segmented into 102812 sentences. A summary of the speech corpora is presented in Table \ref{tab:weaklys_speech_corpora}, whereas the details are presented in our recent work \cite{korzekwa21b_interspeech}.

The speech data are used in all the pronunciation error detection experiments presented in Section \ref{sec:experiments_pronunciation}. From the collected speech, we held out 28 L2 speakers and used them only to assess the performance of the systems in the mispronunciation detection task. It includes 11 Italian and 11 German speakers from the Isle corpus \cite{atwell2003isle}, and 6 Polish speakers from the GUT Isle corpus \cite{Weber2020}. The human speech training data is extended with synthetic pronunciation errors generated by the methods presented in Section \ref{sec:synth_speech_generation_methods}. 

\begin{table}
\tbl{Summary of human speech corpora used in the pronunciation error detection experiments. * - audiobooks read by volunteers from all over the world \cite{Zen2019} }
{
  \centering
  \begin{tabular}{lll}
    \toprule  
   Native Language & Hours & Speakers \\
    \midrule
    English & 90.47 & 640\\ 
    Unknown* & 19.91 & 285\\  
    German and Italian & 13.41 & 46\\  
    Polish & 1.49 & 12\\ 
    \bottomrule
  \end{tabular}
  }
    \label{tab:weaklys_speech_corpora}
\end{table}

\subsection{Corpora of isolated words}
\label{sec:speech_corpora_words}

The speech corpora consist of human and synthetic speech. The data were divided into training and testing sets, with separate speakers assigned to each set. Human speech includes native (L1) and non-native (L2) English speech. L1 speech corpora are made of TIMIT \cite{garofolo1993darpa} and Arctic \cite{kominek2004cmu}. L2 corpora contain speech from L2-Arctic [32], Porzuczek \cite{porzuczek2017english}, and our own recordings of 25 speakers (23 Polish, 1 Ukrainian and 1 Lithuanian). The synthetic data were generated using the T2S method and are only included in the training set. The data are summarized in Table \ref{tab:lexical_stress_speech_corpora_traintest}. For a more detailed description of speech corpora, see Section 4 of our recent work \cite{korzekwa21_interspeech}. The speech corpora of isolated words are used in the lexical stress error detection experiment presented in Section \ref{sec:experiments_lexical_stress}.

\begin{table}
 \tbl{Details of the training and test sets for the lexical stress error detection model.}
 {
\begin{tabular}{llll}
    \toprule  
    Data set  & Speakers  (L2) & Words (unique) &  Stress Errors \\
    \midrule
    Train set (human) & 473 (10) & 8223 (1528) & 425\\
    Train set (TTS) & 1 (0) & 3937 (1983) & 2005\\
    Test set (human) & 176 (21) & 2108 (378) & 189\\
    \bottomrule
  \end{tabular}
 }
  \label{tab:lexical_stress_speech_corpora_traintest}
\end{table}

\section{Experiments}
\label{sec:experiments_pronunciation}

\subsection{Generation of mispronounced speech}
\label{sec:p2p_gen_of_incorrectly_pron_speech}

\subsubsection{Experimental setup}

The effect of using synthetic pronunciation errors based on the P2P, T2S and S2S methods is evaluated in the task of detecting pronunciation errors in spoken sentences at the word level. First, we analyze the P2P method by comparing it with the state-of-the-art techniques and measure the effect of adding synthetic pronunciation errors to the training data. We then compare P2P with T2S and S2S to assess the benefits of using more complex methods of generating pronunciation errors. The accuracy of detecting pronunciation errors is reported in standard Area Under the Curve (AUC), precision and recall metrics. 

\subsubsection{Overview of our WEAKLY-S model}

We use the pronunciation error detection model (WEAKLY-S) recently proposed by us \cite{korzekwa21b_interspeech}. To train the model, the human speech training set is extended with 292,242 utterances of L1 speech with synthetically generated pronunciation errors. To generate pronunciation errors, the P2P, T2S, and S2S methods described in Section \ref{sec:synth_speech_generation_methods} are used.

The WEAKLY-S model produces probabilities of mispronunciation for all words, conditioned by the spoken sentence and canonical phonemes. Mispronunciation errors include phoneme replacement, addition, deletion, or an unknown speech sound. During training, the model is weakly supervised, in the sense that only mispronounced words in L2 speech are marked by listeners and the data do not have to be phonetically transcribed. Due to the limited availability of L2 speech and the fact that it is not phonetically transcribed, the model is more likely to overfit. To solve this problem, the model is trained in a multi-task setup. In addition to the primary task of detecting mispronunciation error at the word level, the second task uses a phoneme recognizer which is trained on automatically transcribed L1 speech. Both tasks share components of the model, which makes the primary task less likely to overfit.

\begin{figure}
\centering
\includegraphics[height=2.9cm]{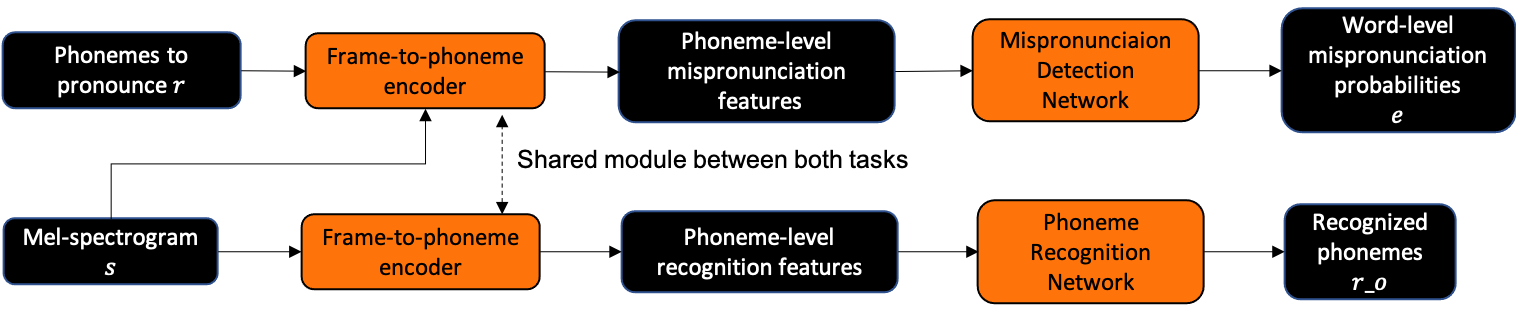}
\caption{Architecture of the WEAKLY-S model for word-level pronunciation error detection trained in the multi-task setup. Task 1 - to detect pronunciation errors $e$. Task 2 - to recognize phonemes $r_o$.} 
\label{fig:weakly_architecture}
\end{figure}

The architecture of the pronunciation error detection model is shown in Figure \ref{fig:weakly_architecture}. The model consists of two sub-networks. The Mispronunciations Detection Network (MDN) detects word-level pronunciation errors $\mathbf{e}$ from the audio signal $\mathbf{s}$ and canonical phonemes $\mathbf{r}$, while the Phoneme Recognition Network (PRN) recognizes phonemes $\mathbf{r_o}$ pronounced by a speaker from the audio signal $\mathbf{s}$. The detailed model architecture is presented in Section 2 of our recent work \cite{korzekwa21b_interspeech}.

\subsubsection{Results - P2P method}

We conducted an ablation study to measure the effect of removing synthetic pronunciation errors from the training data. We trained four variants of the WEAKLY-S model to measure the effect of using synthetic data against other elements of the model. WEAKLY-S is a complete model that also includes synthetic data during training. In the NO-SYNTH-ERR model, we exclude synthetic samples of mispronounced L1 speech, significantly reducing the number of mispronounced words seen during training from 1,129,839 to just 5,273 L2 words. The NO-L2-ADAPT variant does not fine-tune the model on L2 speech, although it is still exposed to L2 speech while being trained on a combined corpus of L1 and L2 speech. The NO-L1L2-TRAIN model is not trained on L1/L2 speech, and fine-tuning on L2 speech starts from scratch. This means that this model will not use a large amount of phonetically transcribed L1 speech data and ultimately no secondary phoneme recognition task will be used. 

\begin{figure}
\centering
\includegraphics[height=4.4cm]{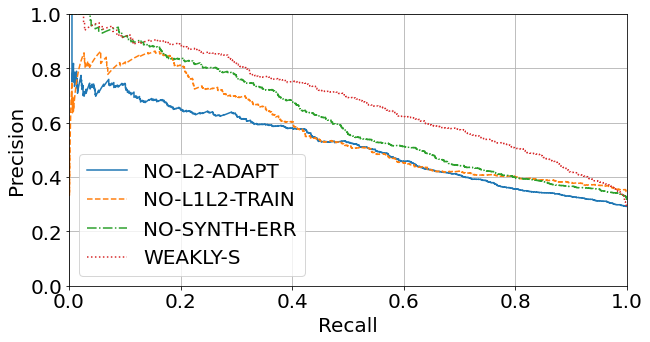}
\caption{Precision-recall curve for the ablation study on the GUT Isle corpus, illustrating the effect of using synthetic pronunciation errors generated by the P2P method.} 
\label{fig:weaklys_ablation_precision_recall_plots}
\end{figure}

L2 fine-tuning (NO-L2-ADAPT) is the most important factor influencing the performance of the model (Fig. \ref{fig:weaklys_ablation_precision_recall_plots} and Table \ref{tab:weaklys_ablation_study}), with an AUC of 0.517 compared to 0.686 for the full model. Training the model on both L2 and L1 human speech together is not enough. This is because L2 speech accounts for less than 1\% of the training data and the model naturally leans towards L1 speech. The second most important feature is training the model on a combined set of L1 and L2 speech (NO-L1L2-TRAIN), with an AUC of 0.565. L1 speech accounts for over 99\% of training data. These data are also phonetically transcribed, and therefore can be used for the phoneme recognition task. The phoneme recognition task acts as a 'backbone' and reduces the effect of overfitting in the main task of detecting errors in the pronunciation of words. Finally, excluding synthetically generated pronunciation errors (NO-SYNTH-ERR) reduces an AUC from 0.686 to 0.615. Although, the synthetic data provides the least improvement to the model, it still increases the accuracy of the model by 11.5\% in AUC, contributing to setting up a new state-of-the-art. 

\begin{table}
  \tbl{Ablation study for the GUT Isle corpus to show the effect of using synthetic data and other elements of the WEAKLY-S model.}
  {
  \centering
   \begin{tabular}{lllll}
    \toprule
     Model & Description & AUC & Precision [\%] & Recall [\%] \\
    \midrule
     NO-L2-ADAPT & No fine-tuning on L2 speech & 0.517 & 57.89 & 40.11 \\ 
     NO-L1L2-TRAIN & No pretraining on L1\&L2 speech & 0.565 & 59.73 & 40.20 \\  
     NO-SYNTH-ERR & \makecell[l]{ No synthetically generated pronunciation \\ errors in the training data}& 0.615 & 67.22 & 40.38 \\  
     WEAKLY-S & Complete model & \textbf{0.686} & 75.25  & 40.38 \\    
    \bottomrule
  \end{tabular}
  }
\label{tab:weaklys_ablation_study}
\end{table}

We compare the WEAKLY-S model with two state-of-the-art baselines. The Phoneme Recognizer (PR) model by Leung et al. \cite{leung2019cnn} is our first baseline. The PR is based on the CTC loss \cite{graves2012connectionist} and outperforms multiple alternative approaches of pronunciation assessment. The original CTC-based model uses a hard likelihood threshold applied to the recognized phonemes. To compare it with two other models, following our recent work \cite{korzekwa2021mispronunciation}, we have replaced the hard likelihood threshold with a soft threshold. The second baseline is  PR extended by the pronunciation model (PR-PM model \cite{korzekwa2021mispronunciation}). The pronunciation model takes into account the phonetic variability of the speech spoken by native speakers, which results in greater precision in detecting pronunciation errors. The results are shown in Table \ref{tab:weaklys_accuracy_metrics}. It turns out that the WEAKLY-S model outperforms the second-best model in terms of an AUC by 30\% from 0.528 to 0.686 and precision by 23\% from 0.612 to 0.752 on the GUT Isle Corpus of Polish speakers. We are seeing similar improvements on the Isle Corpus of German and Italian speakers. The use of synthetic data is an important contribution to the performance of the WEAKLY-S model.

\begin{table}
  \tbl{Accuracy metrics of detecting word-level pronunciation errors. WEAKLY-S vs. baseline models.}
  {
  \centering
   \begin{tabular}{llll}
    \toprule
     Model & AUC & Precision [\%,95\%CI] & Recall [\%,95\%CI] \\
    \midrule
    \multicolumn{4}{c}{\textbf{Isle corpus (German and Italian)}} \\
    PR & 0.555 & 49.39 (47.59-51.19) & 40.20 (38.62-41.81)\\ 
    PR-PM & 0.480 & 54.20 (52.32-56.08) & 40.20 (38.62-41.81)\\  
    WEAKLY-S & \textbf{0.678} & 71.94 (69.96, 73.87) & 40.14 (38.56, 41.75) \\  
	
	 \multicolumn{4}{c}{\textbf{GUT Isle corpus (Polish)}} \\
     PR & 0.528 & 54.91 (50.53-59.24) & 40.29 (36.66-44.02)\\ 
     PR-PM & 0.505 & 61.21 (56.63-65.65) & 40.15 (36.51-43.87)\\  
     WEAKLY-S & \textbf{0.686} & 75.25 (71.67-78.59) & 40.38 (37.52-43.29)\\    
    \bottomrule
  \end{tabular}
  }
   \label{tab:weaklys_accuracy_metrics}
\end{table}

\subsubsection{Results - T2S and S2S methods}
\label{sec:s2s_experiment}
The main limitation of the P2P method is that it does not generate a new speech signal. The method introduces mispronunciations by operating only on the sequence of phonemes for the corresponding speech. In this experiment, we demonstrate the T2S and S2S methods that can directly generate a speech signal to overcome this limitation. The S2S method introduces mispronunciations into the input native speech while preserving the prosody (phoneme durations) and timbre of the voice. Preserving speech attributes other than pronunciation increases speech variability during training and makes the pronunciation error detection model more reliable during testing. The T2S method can be considered as a simplified variant of the S2S method, in which there is only text as input. 

The T2S and S2S methods are compared with the P2P method. Three WEAKLY-S models are trained, differing in the technique of generating mispronounced speech contained in the training data. The S2S method outperforms the P2P method by increasing an AUC score by 9\% from 0.686 to 0.749 in the Gut Isle corpus of Polish speakers (Table \ref{tab:speech_to_speech_auc_results}). Additionally, an AUC increases from 0.815 to 0.834 for major pronunciation errors (Table \ref{tab:speech_to_speech_auc_results_high_sev_real_speech}), according to a similar experiment presented in Section 3.4 of \cite{korzekwa21b_interspeech}. Interestingly, the T2S method is only slightly better than the P2P method, which suggests that the variability of the generated mispronounced speech provided by the S2S method is really important. The presented experiments show the potential of the S2S method in improving the accuracy of detecting pronunciation errors. The S2S method is able to control voice timbre, phoneme duration, and pronunciation, opening the door to transplanting all three properties from non-native speech and potentially further improving the accuracy of the model. 

One downside of the S2S method is its complexity. Compared to the straightforward P2P method, the 9\% improvement in an AUC is associated with high costs. The method involves training a complex multi-speaker S2S model to convert between input and output mel-spectrograms and requires training a Universal Vocoder model to convert a mel-spectrogram into a raw speech signal. 

To better understand what prevents the model from achieving higher accuracy, we measure the performance of the model on synthetic pronunciation errors. We divide all synthetic pronunciation errors into four categories to reflect the severity of pronunciation errors. The `low' category includes mispronounced words with only one mismatched phoneme between the canonical and pronounced phonemes of the word. The `medium' category  includes two mispronounced phonemes. The `high' category gets three, and the 'very high' category includes four mispronounced errors. The AUC across different severity levels varies from 0.928 (low severity) to 1.00 (very high severity) as shown in Table \ref{tab:speech_to_speech_auc_synth_speech}. These AUC values are significantly higher than the results for non-native human  speech, suggesting that making synthetic speech errors more similar to non-native speech may improve the accuracy of detecting pronunciation errors.

\begin{table}
  \tbl{Comparison of the P2P, T2S and S2S methods in the task of pronunciation error detection assessed on the GUT Isle corpus.}
  {
  \centering
   \begin{tabular}{llll}
    \toprule
     Model & AUC & Precision [\%] & Recall [\%] \\
    \midrule
     P2P & 0.686 &  75.25 (71.67-78.59) & 40.38 (37.52-43.29) \\   
     T2S & 0.695 & 76.15 (72.59-79.36) & 40.25 (37.44-43.22) \\ 
     S2S & 0.749 & 80.45 (76.94-83.47) & 40.12 (37.12-43.02) \\  
    \bottomrule
  \end{tabular}
  }
  \label{tab:speech_to_speech_auc_results}
\end{table}

\begin{table}
  \tbl{Comparison of the P2P, T2S and S2S methods in the task of pronunciation error detection assessed on the GUT Isle corpus only for major pronunciation errors.}
  {
  \centering
   \begin{tabular}{llll}
    \toprule
     Model & AUC & Precision [\%] & Recall [\%] \\
    \midrule
     P2P & 0.815 &  91.67 (88.55-94.45)   & 40.31 (37.43-43.23) \\   
     T2S & 0.819 & 92.11 (89.09-94.83) & 40.21 (36.81-43.31)\\  
     S2S & 0.834 & 93.54 (90.53-96.23)  & 40.15 (37.26-43.11)\\  
    \bottomrule
  \end{tabular}
  }
  \label{tab:speech_to_speech_auc_results_high_sev_real_speech}
\end{table}

\begin{table}
  \tbl{Accuracy (AUC) in  detecting  pronunciation errors assessed in synthetic speech at different severity levels of mispronunciation for the best S2S method.}
  {
  \centering
   \begin{tabular}{ll}
    \toprule
     Severity & AUC \\
    \midrule
    Low (phoneme distance=1) & 0.928\\   
    Medium (phoneme distance=2) & 0.974 \\  
    High (phoneme distance=3) & 0.993 \\  
    Very High (phoneme distance=4) & 1.00 \\ 
    \bottomrule
  \end{tabular}
  }
  \label{tab:speech_to_speech_auc_synth_speech}
\end{table}

\subsection{Model of native speech pronunciation}

\subsubsection{Experimental setup}

The P2P, T2S, and S2S are generative models that provide the probability of generating a particular output sequence. This probability can be used directly to detect pronunciation errors without generating the mispronounced speech and adding it to the training data. In this experiment, we show how to apply this approach in practice.

One of the challenges in detecting pronunciation errors is that a native speaker can pronounce a sentence correctly in many ways. The classic approach for detecting pronunciation errors is based on identifying the difference between pronounced and canonical phonemes. All pronunciations that do not correspond precisely to the canonical pronunciation will result in false pronunciation errors. One way to solve this problem is to use the P2P technique to create a native speech Pronunciation Model (PM) that determines the probability that a sentence is pronounced by a native speaker. A low likelihood value indicates a high probability of mispronunciation. 

To evaluate the performance of the PM model, the pronunciation error detection model has been designed such that the PM model can be turned on and off. To disable the PM, we are modifying  it so that it only takes into account  one way of correctly pronouncing a sentence. In an ablation study, we measure  whether the PM model improves the accuracy in detecting pronunciation errors at the word level. Note that in this experiment, synthetically generated pronunciation errors are not used explicitly. Instead, the native speech pronunciation model is used to implicitly represent the generative speech process. 

\subsubsection{Overview of the pronunciation error detection model}

\begin{figure}
\centering
\includegraphics[height=2.1cm]{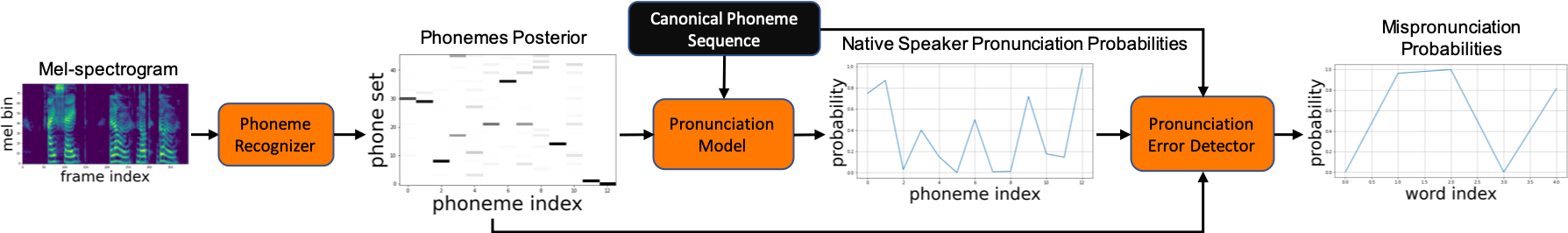}
\caption{Architecture of the system for detecting mispronounced words in a spoken sentence based on the native speech pronunciation model.} 
\label{fig:uncertainty_modeling_architecture}
\end{figure}

The design of the pronunciation error detection model consists of three subsystems: a Phoneme Recognizer (PR), a Pronunciation Model (PM), and a Pronunciation Error Detector (PED), shown in Figure \ref{fig:uncertainty_modeling_architecture}. First, the PR model estimates a belief over the phonemes produced by the student, intuitively representing the uncertainty in the student's pronunciation. The PM model transforms this belief into a probability that a native speaker would pronounce the sentence this way, given the phonetic variability. Finally, the PED model decides which words were mispronounced in the sentence by processing three pieces of information: a) what the student pronounced, b) how likely it is that the native speaker would pronounce it that way, and c) what the student was supposed to pronounce. Details of the entire model of pronunciation error detection are presented in Section 3 of our recent work \cite{korzekwa2021mispronunciation}. We will now only show the details of the PM model that are relevant to this experiment. 

\subsubsection{Overview of the native speech pronunciation model}

PM is an encoder-decoder neural network, following Sutskever et al. \cite{sutskever2014sequence}. Instead of building a text-to-text translation system between two languages, we use it for the P2P conversion. The sequence of phonemes $\mathbf{r}$ that the native speaker was supposed to pronounce is converted to the sequence of phonemes  $\mathbf{r^{'}}$ they had pronounced, denoted as  $\mathbf{r^{'}} \sim p(\mathbf{r^{'}}|\mathbf{r})$. Once trained,  PM acts as a probability mass function, computing the probability sequence $\boldsymbol{\pi}$ of the recognized phonemes $\mathbf{r_o}$ pronounced by the student conditioned by the expected (canonical) phonemes $\mathbf{r}$.  PM is denoted as in Eq. \ref{eq_proposed_model}.

\begin{equation}
\boldsymbol{\pi}=\sum_{\mathbf{r_o}} p(\mathbf{r_o}|\mathbf{o})p(\mathbf{r^{'}}=\mathbf{r_o}|\mathbf{r})
\label{eq_proposed_model}
\end{equation}
The PM model is trained on P2P speech data generated automatically by passing the speech of the native speakers through the PR. By using PR to annotate the data, we can make the PM model more robust against possible phoneme recognition inaccuracies in  PR at the time of testing.

\subsubsection{Results}

The complete model with PM enabled is called  PR-PM that stands for a Phoneme Recognizer + Pronunciation Model. The model with PM turned off is called PR-LIK that stands for Phoneme Recognizer outputting the likelihoods of recognized phonemes. PR-LIK is an extension of the PR-NOLIK model --  the mispronunciation detection model proposed by Leung et al. \cite{leung2019cnn} that only returns the most likely recognized phonemes and does not use phoneme likelihoods to detect pronunciation errors.  PR-NOLIK detects mispronounced words based on the difference between the canonical and recognized phonemes. Therefore, this system does not offer any flexibility in optimizing the model for higher precision by fine-tuning the threshold applied to the phoneme recognition probabilities.

\begin{figure}
\centering
\includegraphics[height=4.4cm]{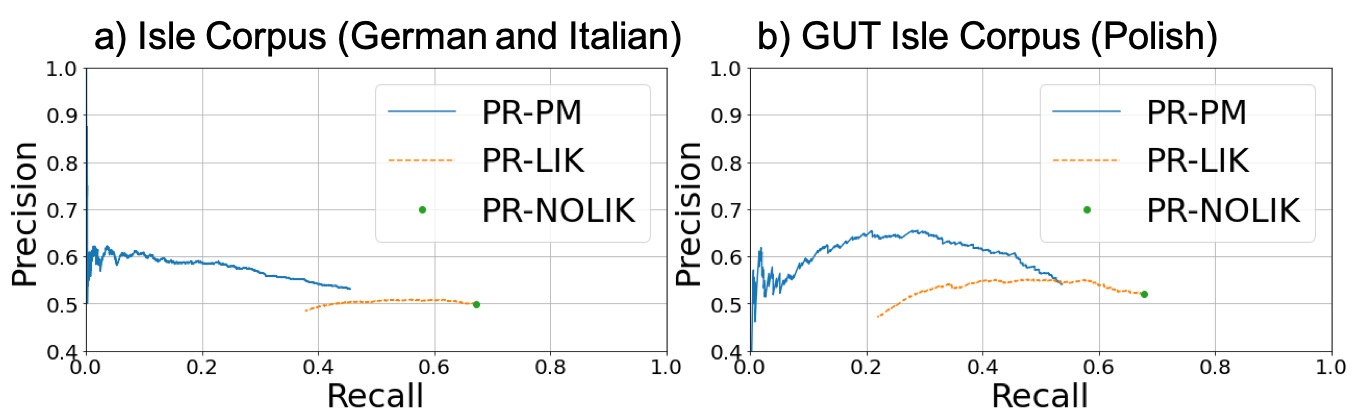}
\caption{Precision-recall curves for the evaluated systems to measure the effect of using the PM model in detecting pronunciation errors. PR-PM - full model with the PM enabled. PR-LIK - the PR-PM model with the PM disabled. PR-NOLIK - non-probabilistic variant of the PR-LIK  model proposed by Leung et al. \cite{leung2019cnn}.} 
\label{fig:pr_uncertainty_modeling_precision_recall_plot}
\end{figure}

Turning off PM reduces the precision between 11\% and 18\%, depending on the decrease in recall between 20\% to 40\%,  as shown in Figure \ref{fig:pr_uncertainty_modeling_precision_recall_plot}. One example where the PM helps is the word `enough' that can be pronounced in two similar ways: /ih n ah f/ or /ax n ah f/ (short `i' or `schwa' phoneme at the beginning.) The PM can take into account the phonetic variability and recognize both versions as correctly pronounced. Another example is coarticulation \cite{hieke1984linking}. Native speakers tend to merge phonemes of adjacent words. For example, in the text `her arrange' /hh er - er ey n jh/, two adjacent phonemes /er/ can be pronounced as one phoneme: /hh er ey n jh/. The PM model can correctly recognize multiple variations of such pronunciations.

Complementary to the precision-recall curve shown in Figure \ref{fig:pr_uncertainty_modeling_precision_recall_plot}, we present in Table \ref{tab:pr_pm_results} one configuration of the precision and recall scores for the PR-LIK and PR-PM systems. This configuration is chosen in a way to: a) make the recall for both systems close to the same value, and b) to illustrate that the PR-PM model has much greater potential to increase precision than the PR-LIK system. A similar conclusion can be drawn by checking various different precision and recall configurations in the precision and recall plots for both Isle and GUT Isle corpora.

\begin{table}
\tbl{Precision and recall of detecting word-level mispronunciations. CI - Confidence Interval. PR-PM - full model with the PM enabled. PR-LIK - the PR-PM model with the PM disabled.}
  {
  \centering
  \begin{tabular}{lll}
    \toprule  
   Model  & Precision [\%,95\%CI] & Recall [\%,95\%CI] \\
    \midrule
    \multicolumn{3}{c}{\textbf{Isle corpus (German and Italian)}} \\
    PR-LIK & 49.39 (47.59-51.19) & 40.20 (38.62-41.81)\\ 
    PR-PM & 54.20 (52.32-56.08) & 40.20 (38.62-41.81)\\  
     \multicolumn{3}{c}{\textbf{GUT Isle corpus (Polish)}} \\
    PR-LIK & 54.91 (50.53-59.24) & 40.29 (36.66-44.02)\\ 
    PR-PM & 61.21 (56.63-65.65) & 40.15 (36.51-43.87)\\  
    \bottomrule
  \end{tabular}
  }
  \label{tab:pr_pm_results}
\end{table}

\subsection{Lexical stress error detection}
\label{sec:experiments_lexical_stress}

\subsubsection{Experimental setup}

The full CAPT learning experience includes both the detection of pronunciation and lexical stress errors. To investigate the potential of speech generation in the lexical stress error detection task, we evaluate the T2S method, which is a simpler version of the S2S method evaluated in Section \ref{sec:s2s_experiment}. 

The lexical stress error detection model is trained to measure the benefits of employing synthetic mispronounced speech. The first model, denoted as Att\_TTS is based on an attention mechanism and is trained on both human and synthetic speech with pronunciation errors. In this model, 1980 the most popular English words \cite{michel2011quantitative} were synthesized with correct and incorrect stress patterns using the method outlined in Section \ref{sec:t2s_method}, and added to the speech corpora of isolated words presented in  Section \ref{sec:speech_corpora_words}. The Att\_NoTTS model is trained only on human speech. Each of the two models presented has its simpler version without the attention mechanism, marked as NoAtt\_TTS and NoAtt\_NoTTS. Both models will help to understand whether the benefits of using synthetic pronunciation errors depend on the model capacity.

The accuracy of detecting lexical stress errors is measured in terms of an AUC metric. To be comparable to the study by Ferrer et al. \cite{ferrer2015classification}, we use precision as an additional metric, while setting recall to 50\%.

\subsubsection{Overview of the lexical stress detection model}

\begin{figure}
\centering
\includegraphics[height=3.3cm]{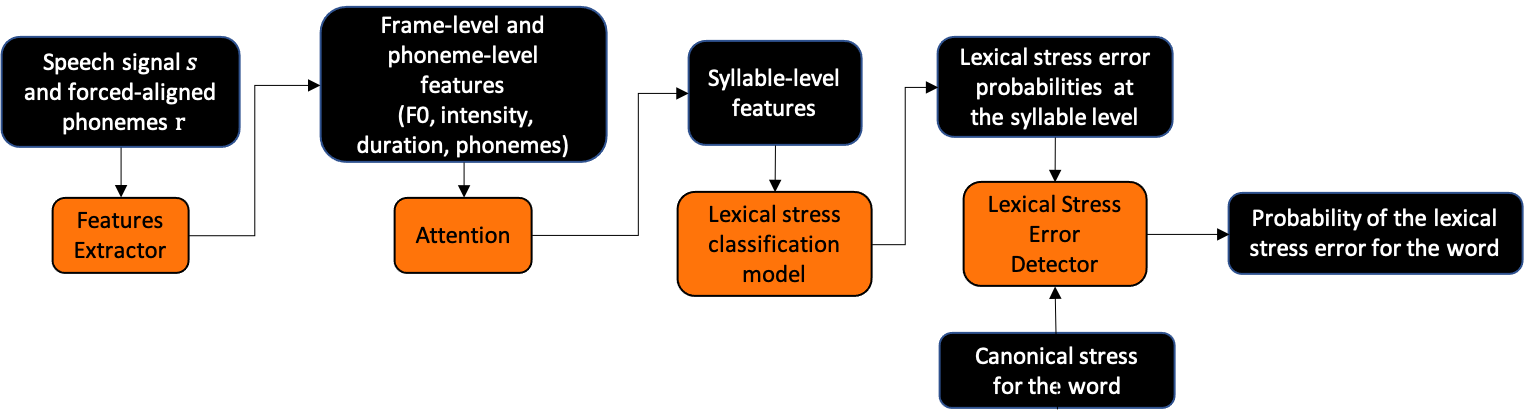}
\caption{Attention-based model for the detection of lexical stress errors.} 
\label{fig:lexical_stress_architecture}
\end{figure}

As shown in Figure \ref{fig:lexical_stress_architecture}, the lexical stress error detection model consists of three subsystems: Feature Extractor, Attention-based Classification Model, and Lexical Stress Error Detector. The Feature Extractor extracts prosodic features and phonemes from the speech signal $\mathbf{s}$ and the forced-aligned canonical phonemes $\mathbf{r}$. Prosodic features include: F0, intensity [dB SPL] and duration of phonemes. The F0 and intensity features are computed at the frame level. The Attention-based Classification Model uses the attention mechanism \cite{vaswani2017attention} to map frame-level and phoneme-level features to a syllable-level representation. It then produces lexical stress error probabilities at the syllable level. The Lexical Stress Error Detector reports a lexical stress error if the expected (canonical) and estimated lexical stress for a given syllable do not match and the corresponding probability is higher than the specified threshold. The detailed architecture of the model is presented in Section 3 of our recent work \cite{korzekwa21_interspeech}.

The NoAtt\_TTS and NoAtt\_NoTTS models do not have the attention mechanism. Instead, as a representation at the syllable level, they use the average acoustic feature values for the corresponding syllable nucleus. The hypothesis is that synthetic data will not be beneficial to a simpler model due to its limited capacity. 

\subsubsection{Results}

\begin{figure}
\centering
\includegraphics[height=4.4cm]{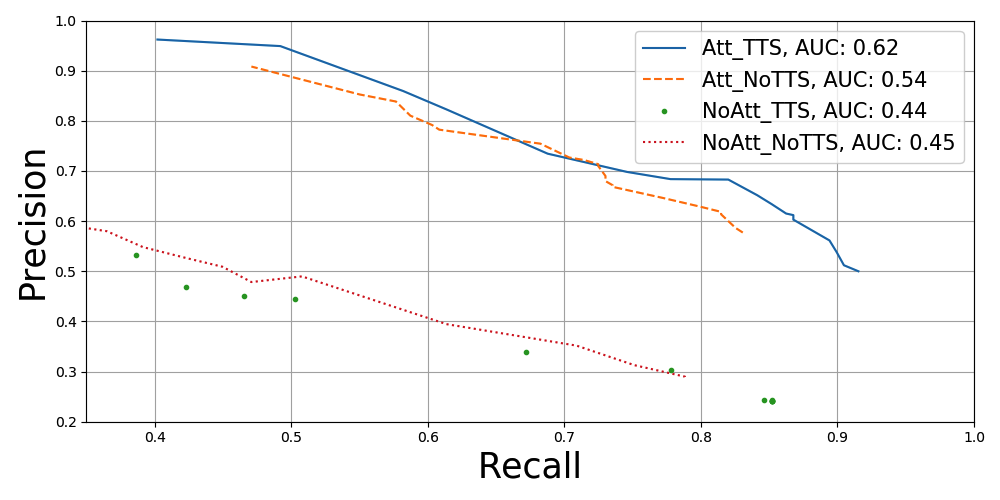}
\caption{Precision-recall curves for lexical stress error detection models.} 
\label{fig:lexical_stress_error_detection_precision_recall_plot}
\end{figure}

Enriching the training set with the incorrectly stressed words increases an AUC score from 0.54 to 0.62 (Att\_TTS vs. Att\_NoTTS in Figure \ref{fig:lexical_stress_error_detection_precision_recall_plot} and Table \ref{tab:lexical_stress_precision_recall_acc}). Data augmentation helps because it increases the number of words with incorrect stress patterns in the training set. This prevents the model from using the strong correlation between phonemes and lexical stress in the correctly stressed words. Using data augmentation in the simpler model without the attention mechanism slightly reduced an AUC score from 0.45 to 0.44 (NoAtt\_NoTTS vs NoAtt\_TTS). The NoAtt\_TTS model has limited capacity due to not using the attention mechanism to model prosodic features, and thus is unable to benefit from synthetic speech.

We compare our results with the work of Ferrer et al. \cite{ferrer2015classification}. There were 46.4\% (191 out of 411) of incorrectly stressed words in their corpus, well over 9.4\% (189 out of 2109) words in our experiment. The fewer lexical stress errors that users make, the more difficult it is to detect them. Under these conditions, we can state that our lexical stress detection model based on T2S generated synthetic speech achieves higher scores in precision and recall compared to the work of Ferrer et al. \cite{ferrer2015classification}.

\begin{table}
 \tbl{AUC, precision and recall [\%, 95\% Confidence Interval] metrics for lexical stress error detection models.} 
 {
  \centering
  \begin{tabular}{llllll}
    \toprule  
   Model & \makecell[l]{Model with \\attention} & \makecell[l]{Synthetic \\mispronunciations} & AUC & Precision [\%] & Recall[\%]  \\
    \midrule
    Att\_TTS & yes & yes & 0.62 & 94.8 (89.18-98.03) & 49.2 (42.13-56.3) \\
    Att\_NoTTS & yes & no & 0.54 & 87.85 (80.67-93.02) & 49.74 (42.66-56.82)\\
    NoAtt\_TTS & no & yes & 0.44 & 44.39 (37.85-51.09) & 50.26 (43.18-57.34)  \\
    NoAtt\_NoTTS & no & no & 0.45 & 48.98 (42.04-55.95) & 50.79 (43.70-57.86)  \\
    Ferrer et al. \cite{ferrer2015classification} & na & na & na & 95.00 (na-na) & 48.3 (na-na)  \\
    \bottomrule
  \end{tabular}
  }
  \label{tab:lexical_stress_precision_recall_acc}
\end{table}

\section{Conclusions}
\label{sec:conclusions}
 
We propose a new paradigm for detecting pronunciation errors in non-native speech. Rather than focusing on detecting  pronunciation errors directly, we reformulate the detection problem as a speech generation task. This approach is based on the assumption that it is easier to generate speech with specific characteristics than to detect those characteristics in speech with limited availability. In this way, we address one of the main problems of the existing CAPT methods, which is the low availability of mispronounced speech for reliable training of pronunciation error detection models.

We present a unified look at three different speech generation techniques for detecting pronunciation errors based on P2P, T2S and S2S conversion. The P2P, T2S, and S2S methods improve the accuracy of detecting pronunciation and lexical stress errors. The methods outperform strong baseline models and establish a new state-of-the-art. 
The best S2S method outperforms the baseline method \cite{leung2019cnn} by improving the accuracy of detecting pronunciation errors in AUC metric by 41\% from 0.528 to 0.749.  The S2S method has the ability to control many properties of speech, such as voice timbre, prosody (duration), and pronunciation. This opens the door to the generation of mispronounced speech that can mimic certain aspects of non-native speech, such as voice timbre. The S2S method can be seen as a generalization of the simpler methods, T2S and P2P, providing a general framework for building a first-class models of pronunciation assessment. For better reproducibility, in addition to using publicly available speech corpora, we recorded the GUT Isle corpus of non-native English speech  \cite{Weber2020}. The corpus is available to other researchers in the field.

In the future, we plan to extend the S2S method in order to generate synthetic speech as close as possible to non-native speech: a) we will extract the voice timbre from the speech of non-native speakers and transfer it to native speech, following the paper of Merritt et al. on text-free voice conversion \cite{merritt22_icassp_vc}, and b) we will mimic the distribution of pronunciation errors in non-native speech. We expect both changes to increase the accuracy of detecting pronunciation errors in non-native speech. In the long run, we hope to demonstrate that ''synthetic speech is all you need'' by training the model with synthetic speech only and achieving  state-of-the-art results in the pronunciation error detection task. This may revolutionize computer-assisted English L2 learning and CAPT. Moreover, such a paradigm may be transferred to the whole domain of computer-assisted foreign language learning.

\bibliographystyle{tfq}
\bibliography{interacttfqsample}

\end{document}